\documentclass[aps,prl,twocolumn,amsmath,amssymb,showpacs,superscriptaddress]{revtex4-1}

\usepackage{graphicx}
\usepackage{color}
\usepackage{hyperref}
\usepackage{gensymb}

\makeatletter
\@ifundefined{textcolor}{}
{
 \definecolor{BLACK}{gray}{0}
 \definecolor{WHITE}{gray}{1}
 \definecolor{RED}{rgb}{1,0,0}
 \definecolor{GREEN}{rgb}{0,1,0}
 \definecolor{BLUE}{rgb}{0,0,1}
 \definecolor{CYAN}{cmyk}{1,0,0,0}
 \definecolor{MAGENTA}{cmyk}{0,1,0,0}
 \definecolor{YELLOW}{cmyk}{0,0,1,0}
}

\begin{document}

\title{Fragility of Fermi arcs in Dirac semimetals}

\author{Yun Wu}

\author{Na Hyun Jo}
\affiliation{Division of Materials Science and Engineering, Ames Laboratory, Ames, Iowa 50011, USA}
\affiliation{Department of Physics and Astronomy, Iowa State University, Ames, Iowa 50011, USA}

\author{Lin-Lin Wang}
\affiliation{Division of Materials Science and Engineering, Ames Laboratory, Ames, Iowa 50011, USA}

\author{Connor A. Schmidt}

\author{Kathryn M. Neilson}

\author{Benjamin Schrunk}

\author{Przemyslaw Swatek}

\author{Andrew Eaton}

\author{S.~L.~Bud'ko}

\author{P. C. Canfield}
\email[]{canfield@ameslab.gov}

\author{Adam Kaminski}
\email[]{kaminski@ameslab.gov}
\affiliation{Division of Materials Science and Engineering, Ames Laboratory, Ames, Iowa 50011, USA}
\affiliation{Department of Physics and Astronomy, Iowa State University, Ames, Iowa 50011, USA}

\date{\today}

\begin{abstract}
We use tunable, vacuum ultraviolet laser-based angle-resolved photoemission spectroscopy and density functional theory calculations to study the electronic properties of Dirac semimetal candidate cubic PtBi${}_{2}$. In addition to bulk electronic states we also find surface states in PtBi${}_{2}$ which is expected as PtBi${}_{2}$ was theoretical predicated to be a candidate Dirac semimetal. The surface states are also well reproduced from DFT band calculations. Interestingly, the topological surface states form Fermi contours rather than double Fermi arcs that were observed in Na$_3$Bi. The surface bands forming the Fermi contours merge with bulk bands in proximity of the Dirac points projections, as expected. Our data confirms existence of Dirac states in PtBi${}_{2}$ and reveals the fragility of the Fermi arcs in Dirac semimetals. Because the Fermi arcs are not topologically protected in general, they can be deformed into Fermi contours, as proposed by [Kargarian {\it et al.}, PNAS \textbf{113}, 8648 (2016)]. Our results demonstrate validity of this theory in PtBi${}_{2}$.

\end{abstract}


\maketitle

\begin{figure*}[tb]
	\includegraphics[width=6 in]{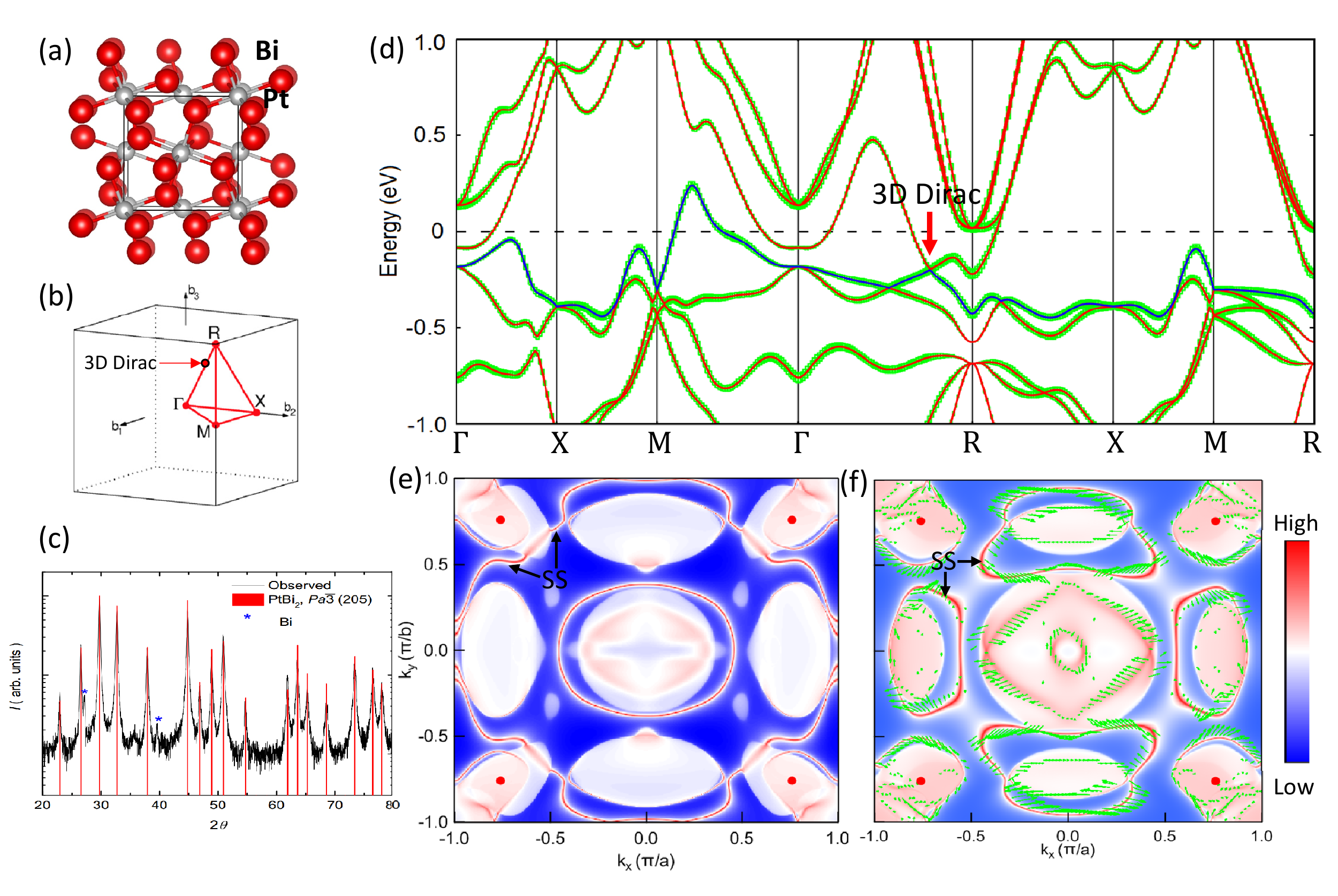}%
	\caption{Crystal structure and calculated band structure of cubic PtBi${}_{2}$.
	(a) Crystal structure of cubic PtBi${}_{2}$ (Pt, white spheres; Bi, red spheres).
	(b) Brillouin zone of PtBi${}_{2}$.
	(c) Powder x-ray diffraction (XRD) pattern of PtBi$_{2}$ (observed pattern, black line; calculated with pyrite structure type [$Pa\bar{3}$, 205], red line; Bi flux peaks, Blue stars.)
	(d) Calculated bulk band structure. The red arrow points to the 3D Dirac point along $\Gamma$-$R$ line. The green color show the magnitude of the projection on Bi $p$ orbitals. It shows the switching of orbital characters at the bulk Dirac point.
	(e) Calculated surface Fermi surface at E${}_{F}$ with surface Green's function using a semi-infinite PtBi${}_{2}$ (001) surface with Bi-termination. 
	(f) Same as (e) but at E${}_{F} + 100~meV$. The red dots in (e) and (f) mark the projections of the 3D Dirac points in (d). The black arrows point to the surface states (SS). The green arrows mark the spin texture of the surface states.
	\label{fig:Fig1}}
\end{figure*}

\begin{figure*}[tb]
	\includegraphics[width=7in]{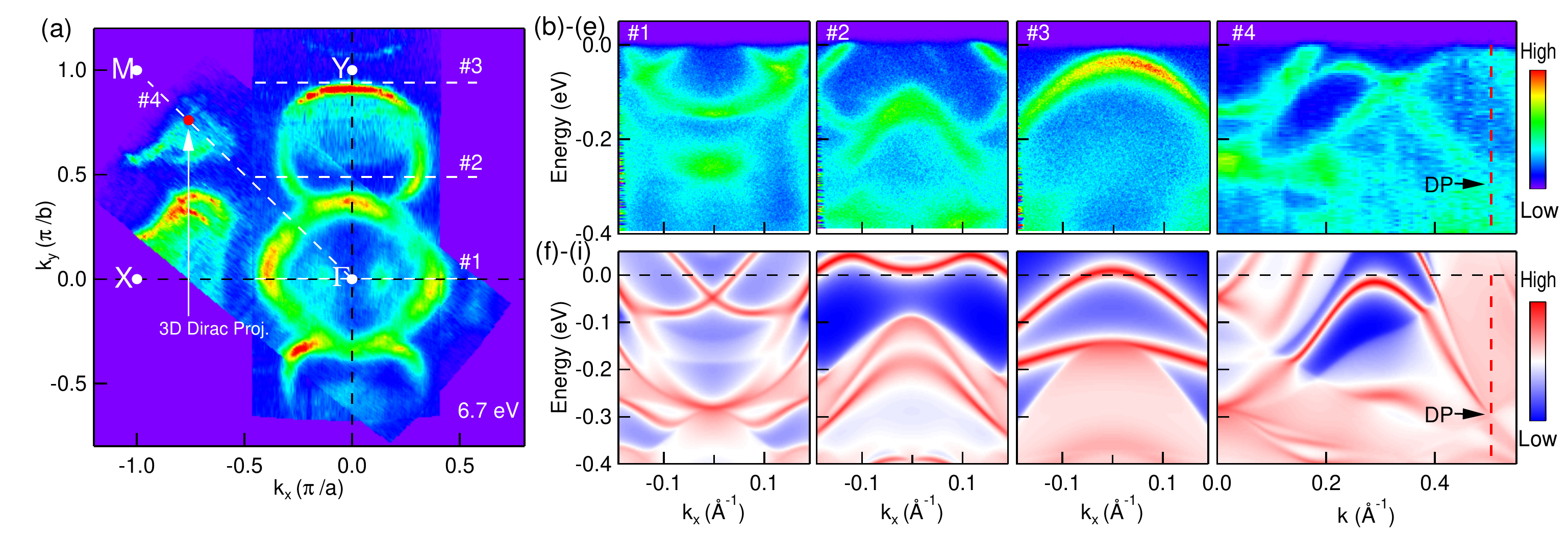}%
	\caption{Fermi surface plot and band dispersions of PtBi${}_{2}$.
	(a) Fermi surface plot - ARPES intensity integrated within 10 meV about the chemical potential. The Fermi surface is generated by overlaying two data sets measured with two different sample orientations. White dots mark high symmetry points; red dot mark the projection of 3D Dirac point.
	(b)--(e) Band dispersions along cuts \#1--\#4 in (a).
	(f)--(i) Calculated dispersions of the surface band along cuts \#1--\#4 in (a). To achieve a better match with (b)--(e), the chemical potentials in (f)--(i) have been shifted upward by $\sim$100~meV. The black arrows and red dashed lines in (e) and (i) mark the location of the 3D Dirac point (DP) according to DFT calculations.
	\label{fig:Fig2}}
\end{figure*}

\begin{figure*}[tb]
	\includegraphics[width=6in]{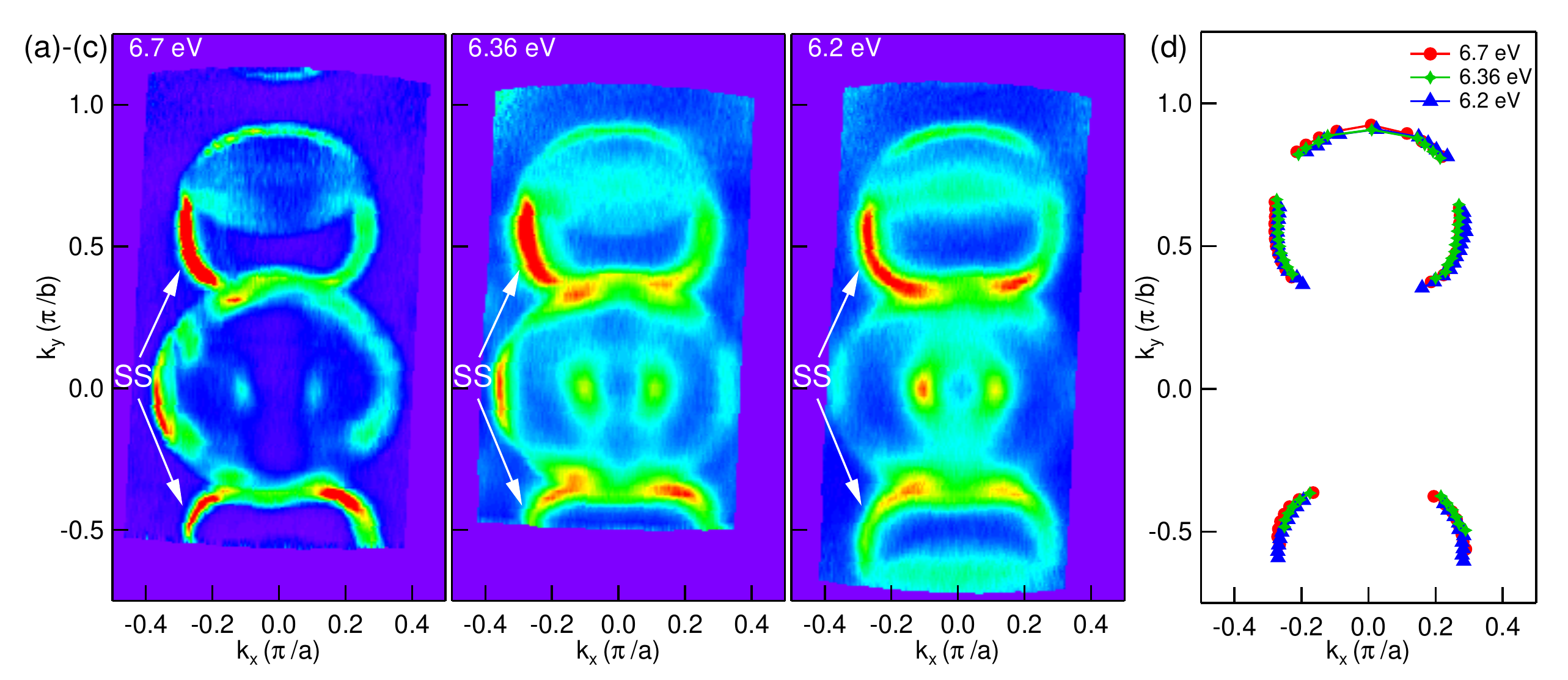}%
	\caption{Fermi surface plots of PtBi${}_{2}$ measured using different photon energies.
	(a)--(c) Fermi surface plots of PtBi${}_{2}$ measured at photon energies of 6.7, 6.36, and 6.05~eV, respectively.
	(d) Fitted locations of high intensity Fermi surface sheets in (a)--(c), showing no obvious photon energy dependence.
	\label{fig:Fig3}}
\end{figure*}

\begin{figure*}[tb]
	\includegraphics[width=7in]{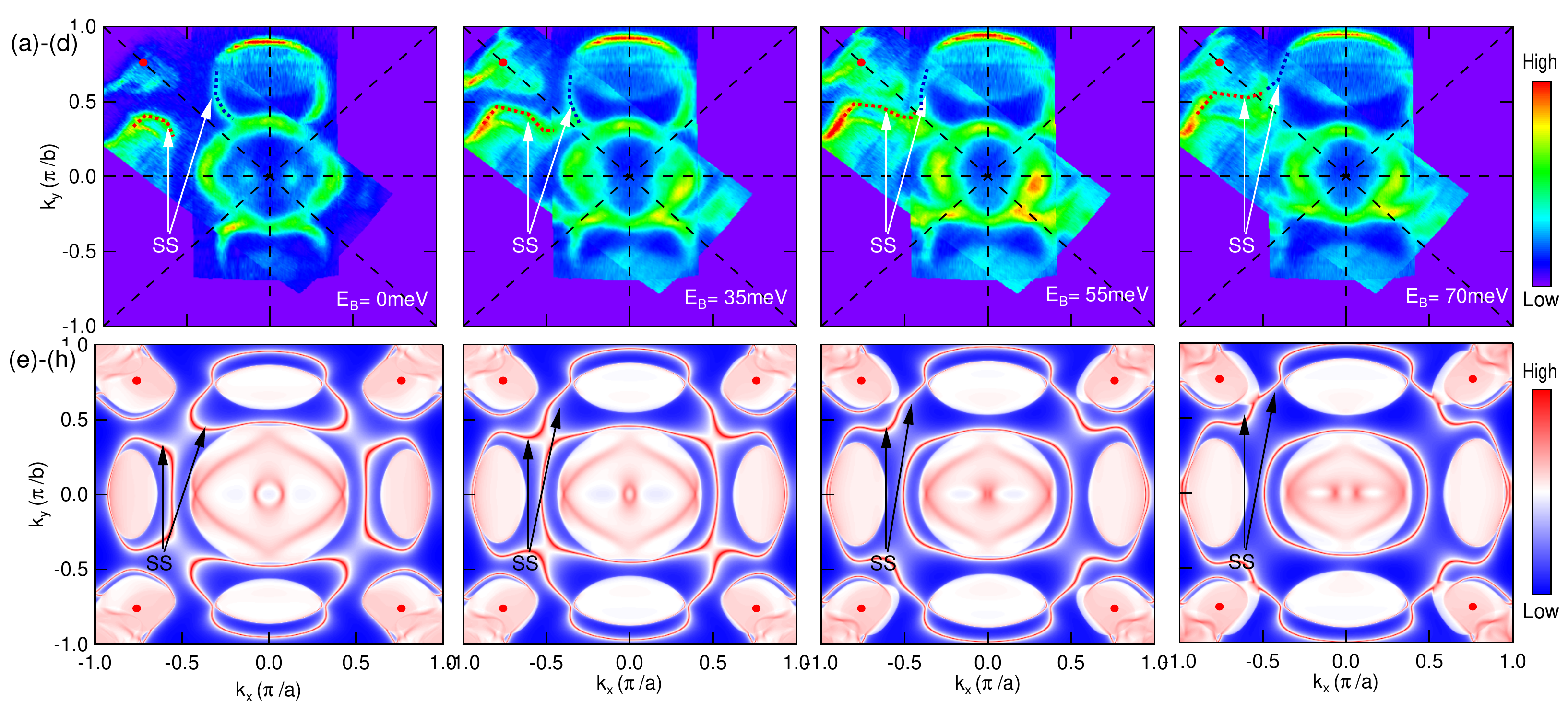}%
	\caption{Constant energy intensity plots of PtBi${}_{2}$.
	(a)--(d) Constant energy intensity plot of PtBi${}_{2}$ at binding energies of 0, 35, 55, and 70~meV. The red dots mark the projections of the 3D Dirac points on the (001) surface. The red and blue dashed lines mark the ``Fermi arc'' surface states (SS) disconnected from or connected with the bulk states containing the projection of the 3D Dirac point.
	(e)--(h) Calculated surface Fermi surface corresponding to (a)--(d), respectively. The white and black arrows in (a)--(h) point to the ``Fermi arc'' SS.
	\label{fig:Fig4}}
\end{figure*}

Topological materials hosting Dirac fermions, Weyl fermions, and Majorana fermions are recognized as having the potential to revolutionize the field of high-performance electronics and fault-tolerant quantum computing~\cite{Nayak2008NonAbelian}. The focus on topological materials started with two-dimensional (2D) Dirac states discovered either in 2D materials such as graphene~\cite{Geim2007Rise} or on the surface of three-dimensional (3D) topological insulators~\cite{Zhang2009Topological, Xia2009Observation, Chen2009Experimental}. Generalizing the 2D Dirac states to the 3D case leads us to the era of 3D Dirac and Weyl semimetals. Na${}_{3}$Bi and Cd${}_{3}$As${}_{2}$ are the two archetypical 3D Dirac semimetals, where the bulk Dirac points are protected by crystal symmetry~\cite{Liu2014Discovery, Neupane2014Observation, Liu2014Stable}. Breaking either the time-reversal~\cite{Wang2016Time} or inversion symmetry~\cite{Xu15SciDis} of a 3D Dirac semimetal can result in a Weyl semimetal with pairs of Weyl nodes that have opposite chirality. The Weyl nodes are monopoles in the Berry curvature with non-zero Chern number that naturally lead to exotic surface states such as Fermi arcs. The discovery of Fermi arcs in both type-I~\cite{Xu15SciDis} and type-II~\cite{Huang2016Spectroscopic, Tamai2016Fermi, Deng2016Experimental} Weyl semimetals confirmed the existence of these exotic states. Although Fermi arc surface states have been predicted~\cite{Wang2012Dirac, Wang2013Three} and observed in 3D Dirac semimetals~\cite{Xu2015Observation, Yi2014Evidence}, the topological nature and stability of these surface states are still under debate~\cite{Kargarian2016Surface, Armitage2018Weyl}. 

Here we demonstrate that the measured surface and bulk band structure of cubic PtBi${}_{2}$ agrees well with calculations and confirms that it is a Dirac semimetal. We further demonstrate that the surface state in this system forms closed Fermi contours rather than double Fermi arcs as reported in Na$_3$Bi \cite{Xu2015Observation}. Since the Fermi arcs in Dirac semimetals are not topologically protected, they can be deformed into closed Fermi contours by a strong surface potential. This was first proposed by Kargarian {\it et al.}~\cite{Kargarian2016Surface} and our data confirms that this indeed occurs in cubic PtBi${}_{2}$. Our results point to fragility of Fermi arcs in Dirac semimetals due to lack of topological protection. 

PtBi${}_{2}$ crystallizes in at least four phases, one of which is a pyrite type with simple cubic crystal structure (space group 205). The pyrite type PtBi${}_{2}$ was predicted to host 3D Dirac points along the $\Gamma$-$R$ line, which are protected by the threefold rotational symmetry~\cite{Gibson2015Three}. Similar to some of the topological materials, such as Dirac node arc metal PtSn${}_{4}$~\cite{Mun2012Magnetic, Wu2016Dirac}, Dirac semimetal Cd${}_{3}$As${}_{2}$~\cite{Liang2015Ultrahigh}, and type-II Weyl semimetal WTe${}_{2}$~\cite{Ali2014Large}, cubic PtBi${}_{2}$ also exhibit extremely large magnetoresistance up to (11.2$\times {10}^{6}$)\% at $T = 1.8$~K in a magnetic field of 33~T~\cite{Gao2017Extremely}. Interestingly, another phase, hexagonal PtBi${}_{2}$ also exhibits giant magnetoresistance~\cite{Yang2016Giant}. Topological surface states have been predicted~\cite{Xu2016Synthesis} and observed~\cite{Yao2016Bulk, Thirupathaiah2018Possible} in hexagonal PtBi${}_{2}$. However, no evidence of the 3D Dirac semimetallic state has been reported in cubic PtBi${}_{2}$ as of yet. 

Single crystals of PtBi${}_{2}$ were grown out of Bi rich binary melts. Elemental Pt and Bi were put into a Canfield Crucible Set~\cite{Canfield2016Use} with initial stoichiometry, Pt${}_{0.05}$Bi${}_{0.95}$, and sealed into an amorphous silica tube. The ampules were heated up to 430\,$\celsius$\, within 5 hours, held for 5 hours, cooled to 300\,\celsius\, over 75 hours, and finally decanted using a centrifuge~\cite{Canfield1992Growth}. Single crystals obtained from the growth were ground to obtain a room temperature powder x-ray diffraction (XRD) pattern with a Rigaku MiniFlex II diffractometer (Cu $K_{\alpha}$ radiation with monochromator).

Band structures with spin-orbit coupling (SOC) in density functional theory (DFT)~\cite{Hohenberg1964Inhomogeneous, Kohn1965Self} have been calculated using a PBE~\cite{Perdew1996Generalized} exchange-correlation functional, a plane-wave basis set and projector augmented wave method~\cite{Blochl1994Projector} as implemented in VASP~\cite{Kresse1996Efficient, Kresse1996Efficiency}. For the bulk band structure of cubic PtBi${}_{2}$, we used the primitive cubic cell of 12 atoms with a Monkhorst-Pack~\cite{Monkhorst1976Special} ($7\times 7\times 7$) $k$-point mesh including the $\Gamma$ point and a kinetic energy cutoff of 230~eV. The convergence with respect to $k$-point mesh was carefully checked, with total energy converged below 1~meV/atom. The experimental lattice parameters have been used with atoms fixed in their bulk positions.  A tight-binding model based on maximally localized Wannier functions~\cite{Marzari1997Maximally, Souza2001Maximally, Marzari2012Maximally} was constructed to reproduce closely the bulk band structure including SOC in the range of ${E}_{F}\pm$1~eV with Pt $sd$ and Bi $p$ orbitals. Subsequently, the Fermi surface and spectral functions of a semi-infinite PtBi${}_{2}$ (001) surface, with Bi-termination, were calculated by using the surface Green’s function methods~\cite{Lee1981SimpleI, Lee1981SimpleII, Sancho1984Quick, Sancho1985Highly} as implemented in WannierTools~\cite{Wu2017WannierTools}. 

Samples used for ARPES measurements were cleaved \textit{in situ} at 40~K under ultrahigh vacuum (UHV). The data were acquired using a tunable VUV laser ARPES system, that consists of a Scienta R8000 electron analyzer, a picosecond Ti:Sapphire oscillator and fourth harmonic generator~\cite{Jiang2014Tunable}. Data were collected with a photon energy of 6.7~eV. Momentum and energy resolutions were set at $\sim$ 0.005~\AA${}^{-1}$ and 2~meV. The size of the photon beam on the sample was $\sim$30~$\mu$m.

Figure 1 shows the crystal structure and calculated electronic structure of PtBi${}_{2}$. Fig.~\ref{fig:Fig1}(a) shows the crystal structure of cubic PtBi${}_{2}$, where the red and white spheres correspond to Bi and Pt atoms, respectively. The acquired XRD patterns are well matched with calculated peaks for pyrite structure type of PtBi$_{2}$ with $Pa\bar{3}$ (205) as shown in Fig.\,\ref{fig:Fig1}(c). The small intensity, extra peaks, marked with blue stars, are associated with residual Bi solvent left on the surface of the crystals. Fig.~\ref{fig:Fig1}(d) presents the bulk electronic structure with the red arrow points to the 3D Dirac point along the $\Gamma$-$R$ line, consistent with the results in Ref.~\onlinecite{Gibson2015Three}. The coordinates of the bulk Dirac points are $\pm$(0.76, 0.76, 0.76)~$\pi/a$ and at ${E}_{F}-197$~meV (Fig.~\ref{fig:Fig1}d). The green shading represents the magnitude of the projection on $p$ orbitals. The constant energy contours calculated with surface Green's function using a semi-infinite PtBi${}_{2}$ (001) surface with Bi-termination are shown in Figs.~\ref{fig:Fig1}(e) and (f). The bulk bands are projected onto a plane and therefore appear as white patches with various degree of red. The surface states are the sharp lines mostly going through the bulk gap regions and sometimes connecting with the bulk states. The green arrows in Fig.~\ref{fig:Fig1}(f) shows the spin texture of the surface states with a helical structure similar to Na${}_{3}$Bi~\cite{Wang2012Dirac}. The red dots are the projections of the 3D Dirac points onto the surface Brillouin zone. Interestingly, at Fermi level (${E}_{F}$), there are distinct Fermi arc surface states connecting the projections of the 3D Dirac point; whereas at 100~meV above ${E}_{F}$, the ``Fermi arc'' surface states break from the bulk projections and form closed loops in between Dirac nodes projections. This calculation result is consistent with the theoretical model in Ref.~\onlinecite{Kargarian2016Surface}, demonstrating the fragility of the ``Fermi arc'' surface states in 3D Dirac semimetals. 

Next, we present detailed ARPES measurements to compare with the results of band structure calculations. Fig.~\ref{fig:Fig2}(a) shows the ARPES intensity map of PtBi${}_{2}$ integrated within 10~meV about the chemical potential. The Fermi surface is generated by overlaying two data sets measured along 0$^\circ$ and $\sim$45$^\circ$ with respect to the crystal $b$ axis. The Fermi surface consists of an electron pocket at the center and several hole pockets along the high symmetry lines. Interestingly, the FS in ARPES resembles the FS calculation at 100~meV above ${E}_{F}$, where the topological surface states are completely disconnected from the projection of the 3D Dirac point (the red solid dot). This chemical potential shift can be better visualized by comparing the band dispersions from ARPES measurements and band structure calculations. The band dispersions along cuts \#1 to \#4 from ARPES and DFT calculations are shown in Figs.~\ref{fig:Fig2}(b)--(e) and (f)--(i), respectively. We can clearly see that the experimental and theoretical results match very well, except that the chemical potential in the calculations needs to be shifted upwards by roughly 100~meV to achieve good agreement. This works luckily to our advantage, saving us the trouble to purposely dope PtBi${}_{2}$ with electrons in order to verify the fragility of the ``Fermi arc'' surface states as shown in Fig.~\ref{fig:Fig1}. Furthermore, by comparing with the calculation results, we can conclude that the high intensity Fermi surface sheets along $\Gamma$-$X$ and $\Gamma$-$Y$ directions in the ARPES data are actually surface states.  

In order to verify that the high intensity Fermi surface sheets indeed have a surface origin, we performed photon energy dependent measurements as shown in Fig.~\ref{fig:Fig3}. Figs.~\ref{fig:Fig3}(a), (b), and (c) show the Fermi surface plots of PtBi${}_{2}$ measured using 6.7, 6.36, and 6.05~eV photons. We can clearly see that the surface states have almost the same intensity profile, whereas the states close to the center of the Brillouin zone vary significantly with different photon energies. To better quantify our observations, we plotted the fitted results of the high intensity locations in Fig.~\ref{fig:Fig3}(d). It is clear that the SS measured from three different photon energies almost perfectly match each other. This photon energy independent characteristic, in tandem with the good match between ARPES and DFT calculations, provide solid arguments for the surface origin of these states.

To demonstrate the fragility of the ``Fermi arc'' surface states in this Dirac semimetal, we plot the constant energy contours at different binding energies in Fig.~\ref{fig:Fig4}. The red dots in Fig.~\ref{fig:Fig4} are the projections of the 3D Dirac points along $\Gamma$-$R$ line. In Fig.~\ref{fig:Fig4}(a), we can clearly see the surface states forming closed loops are well separated from the bulk states containing the projection of the 3D Dirac point (red dot). This can be better visualized in Fig.~\ref{fig:Fig4}(e), the corresponding surface calculations. Moving down in binding energy, we can see the surface state along the $\Gamma$-$Y$ direction starts to detach from the electron pocket surrounding the $\Gamma$ point and moving closer to the other surface states along $\Gamma$-$X$ direction. As we move further down to the binding energy of 55~meV, both surface states along $\Gamma$-$X$ and $\Gamma$-$Y$ directions start to move closer to the bulk states containing the 3D Dirac point projection. Finally, at ${E}_{b}$=70~meV, both surface states merge with the bulk states, matching well the calculation results shown in Fig.~\ref{fig:Fig4}(h). The blue and red dashed lines are guides to the eye marking the path of the surface states. This demonstrates that the topological surface states evolve from closed loops completely dissociated from the projection of 3D Dirac point to actual arc states connecting the bulk states containing the projections. This is a definite experimental proof of the calculation results in Fig.~\ref{fig:Fig1} and the theoretical model in Ref.~\onlinecite{Kargarian2016Surface}. Thus, we show the Fermi arc surface states in this Dirac semimetal are not topological protected and they change from arcs to closed-loops by varying the binding energy. Because there are an even number of (four, to be specific) pairs of Dirac points not located at time reversal invariant momentum locations in PtBi${}_{2}$, it belongs to the ``weak'' Dirac semimetal category, i.e., the surface states observed in PtBi${}_{2}$ could be gapped out by translation symmetry-breaking perturbations~\cite{Kargarian2016Surface}. 

In conclusion, we use ultrahigh resolution ARPES and DFT calculations to demonstrate that the ``Fermi arc'' surface states in 3D Dirac semimetal cubic PtBi${}_{2}$ is not topologically protected, in stark contrast to the Fermi arc surface states in Weyl semimetals. At one binding energy (${E}_{B}=70$~meV), the surface states display an arc form connecting the bulk states containing the projections of the 3D Dirac points. At another binding energy (${E}_{F}$), they become completely disconnected from the projections of the Dirac point and form closed loops in-between the projections at the Fermi level. This demonstrates the fragility of the ``Fermi arc'' surface states in 3D Dirac semimetal cubic PtBi${}_{2}$, consistent with the theoretical model proposed in Ref.~\onlinecite{Kargarian2016Surface}. Furthermore, we observe a Fermi arc-like surface state along $\Gamma$-$X$ direction which is not obvious in DFT calculations. More thorough theoretical understanding is required to better explore the topological nature of this state. 

We would like to thank Mohit Randeria, Yuan-Ming Lu and Peter Orth for very useful comments. Research was supported by the U.S. Department of Energy, Office of Basic Energy Sciences, Division of Materials Science and Engineering. Ames Laboratory is operated for the U.S. Department of Energy by Iowa State University under Contract No. DE-AC02-07CH11358. Y. W. and L. L. W. were supported by Ames Laboratory's Laboratory-Directed Research and Development (LDRD) funding. N.H.J. was supported by the Gordon and Betty Moore Foundation EPiQS Initiative (Grant No. GBMF4411). B. S. was supported by CEM, a NSF MRSEC, under Grant No. DMR-1420451.

Raw data for this manuscript will be available from the authors upon reasonable request.

\bibliography{PtBi2}

\end{document}